\documentclass[10pt,a4paper]{amsart}

\usepackage[normalem]{ulem} 

\usepackage[english]{babel}
\usepackage[utf8]{inputenc}  

\usepackage{epsfig}
\usepackage{float}
\usepackage{subfigure}
\usepackage{moreverb}

\usepackage{xcolor}

\usepackage{afterpage}		
\usepackage{latexsym}
\usepackage{amsmath}
\usepackage{amssymb}
\usepackage{amsfonts}

\usepackage[numbers,sort&compress]{natbib}

\renewcommand{\appendix}{
\setcounter{section}{0}
\renewcommand{\thesection}{\Roman{section}}
\vspace{0.5cm}
{\Large{\bf APPENDIX}}}

\newenvironment{remark}[1][Remark.]{\begin{trivlist}
\item[\hskip \labelsep {\bfseries #1}]}{\end{trivlist}}

\newcommand{\Rb}{\mathbb{R}}

\newcommand{\x}{{\bf x}}

\newcommand{\T}{\mathcal{T}}

\newcommand{\cf}{\mathfrak{c}}

\newcommand{\RR}{\mathcal{R}}

\newcommand{\uu}{{\bf u}}

\newcommand{\vv}{{\bf v}}

\begin{document}

\title[Eulerian flow model]{Refining the Eulerian flow model}


\author{ Magnus Sv\"ard} \footnote{ University of Bergen e-mail: Magnus.Svard@math.uib.no}

\date{\today}

\begin{abstract}
  We revisit the molecular arguments underpinning the Eulerian model for compressible and diffusive flows, and conclude that a heat diffusive term appears to be missing in the original model. By studying a  pure heat transfer problem, we quantify the new term in the updated Eulerian model and evaluate it in the context sound attenuation. Although the new diffusive term is important for certain problems, we also demonstrate that it has a negligible effect on the aerodynamic validation cases that the original model has already successfully passed. Furthermore, the updated Eulerian system is compatible with the weak well-posedness that has previously been established for the original Eulerian system. 
\end{abstract}

\maketitle

\section{Introduction}

The Eulerian flow model (EFM) was proposed in \cite{Svard18} as an alternative to the compressible and heat conductive Navier-Stokes-Fourier equations (NSF).\footnote{The Eulerian model was the original name and in later publications it has also been called the modified or alternative Navier-Stokes equations. Here, we return to the original name as it is a more accurate description.} The new system was motivated by a number of physical problems occurring in the NSF system. These all stem from the Lagrangian mass element derivation of the equations. In contrast, the new model was derived, from the molecular level up to a continuum, in an Eulerian frame and thus circumvented said problems.

The Eulerian system is defined as follows: Let $\x=(x,y,z)$ denote the Cartesian coordinates and $\Omega\subset \Rb^3$ a bounded spatial domain with piecewise smooth boundaries (possibly with holes). The Eulerian flow model is defined by the set of partial differential equations 
\begin{subequations}\label{eulerian} \\
\begin{align}
\partial_t \rho + \nabla\cdot(\rho \vv )&= \nabla\cdot  (\nu \nabla \rho),\label{continuity} \\
\partial_t (\rho \vv) + \nabla\cdot(\rho \vv \otimes \vv) + \nabla p&=\nabla\cdot 
(\nu \nabla \rho \vv)), \qquad t\in [0,\T] \label{momentum} \\
\partial_t (E) + \nabla\cdot(E \vv +p\vv)  &= \nabla\cdot ( \nu \nabla E) + \nabla\cdot(\kappa_T \nabla T), \label{energy} \\ 
p&=\rho R T, \quad \textrm{ideal gas law,}\label{gaslaw}
\end{align}
\end{subequations}
where $\rho,\rho \vv, E$ denote the conserved variables density, momentum and total energy. The velocity vector is, $\vv=(u,v,w)$; $p,T$ are pressure and temperature; the total energy is given by $E=\frac{p}{\gamma-1}+\frac{\rho |\vv|^2}{2}$ where $\gamma=c_p/c_v$ and $c_{p,v}$ are the heat capacities at constant pressure or volume. For ideal gases $1<\gamma\leq 5/3$. The conservative variables are collected in the vector $\uu=(\rho,\rho\vv^T,E)^T$. Furthermore, $R$ is the gas constant and $\nu=\nu(\rho,T)$ is the diffusion coefficient. $\kappa_T$ is the heat diffusion coefficient. In the original model (\cite{Svard18}) it was not included, i.e., $\kappa_T=0$. 

The mathematical structure of the NSF has thus far prevented any existence results. This means that there is no comprehensive theory that aids the design of numerical schemes. The Eulerian system, however, has a different structure and  in \cite{Svard22}, it was proven that the model admits weak entropy solutions with the following diffusion coefficients.
\begin{align}
  \nu &= \frac{\mu_0}{\rho} + \mu_1\rho, \quad \mu_0, \mu_1>0\label{nu}.\\
  \kappa_T&= 4\kappa_rT^3, \quad\kappa_r>0.\label{kappar}
\end{align}
A convergent finite-volume scheme was also derived. 

Furthermore, in \cite{SayyariDalcin21,SayyariDalcin21_2,DolejsiSvard21}, it has been demonstrated that (\ref{eulerian}) produces solutions that are next to indistinguishable from those of the standard Navier-Stokes-Fourier system when $\mu_0$ is taken as the \emph{dynamic viscosity} and $\mu_1=\kappa_r=0$. The addition of $\kappa_r,\mu_1$ was necessary in the existence proof  but since the model is accurate without them, they can  be regarded as technical assumptions. That is, $0<\mu_1<<\mu_0$ and $\kappa_r$ should be very small.   The system (\ref{eulerian}) has also been validated in the incompressible limit in \cite{KajzerPozorski22}.


However, in \cite{SvardMunthe23} it was proposed that the model might miss a heat dissipative mechanism in the energy equation and the molecular origins of such a dissipation term was discussed. In an effort to quantify the potentially missing term, sound attenuation data was used. However, the experimental data did not include measurement accuracy, preventing any definitive conclusions. 

The goal of this paper is to further clarify the presence of and quantify the extra heat dissipative term. To this end, we repeat and expand the molecular arguments for augmenting the model with a heat diffusive term in the energy equation (section \ref{sec:physics}). Turning to its quantification, we note that in \cite{Assael_etal23} it was shown that it is possible to determine the thermal conductivity to a very high accuracy with hot-wire experiments (section \ref{sec:hot_wire}). This allows  for a direct calculation of the missing dissipative term in the Eulerian flow model (section \ref{sec:heat_cond}), and subsequent numerical verifications (section \ref{sec:num_verif}). Furthermore, we relate the new term to the sound attenuation results in \cite{SvardMunthe23} (section \ref{sec:sound}). Since the model (\ref{eulerian}) has already been validated for aerodynamic applications, the extra heat dissipative term should not invalidate these cases. To verify this, we present numerical experiments for a 2-D vortical flow  (section \ref{sec:vortex}) and quantify the differences between the original and updated EFM. Finally, we draw conclusions in section \ref{sec:conclusions}



\section{Molecular dynamics in the Eulerian frame}\label{sec:physics}

In \cite{SvardMunthe23} it was tentatively proposed that a dissipative term is missing in the energy equation of the original EFM. A physical argument for its presence in the energy equation was also provided and we repeat and expand those arguments here.

\begin{figure}[ht]
  \centering
  \subfigure{\includegraphics[width=8cm]{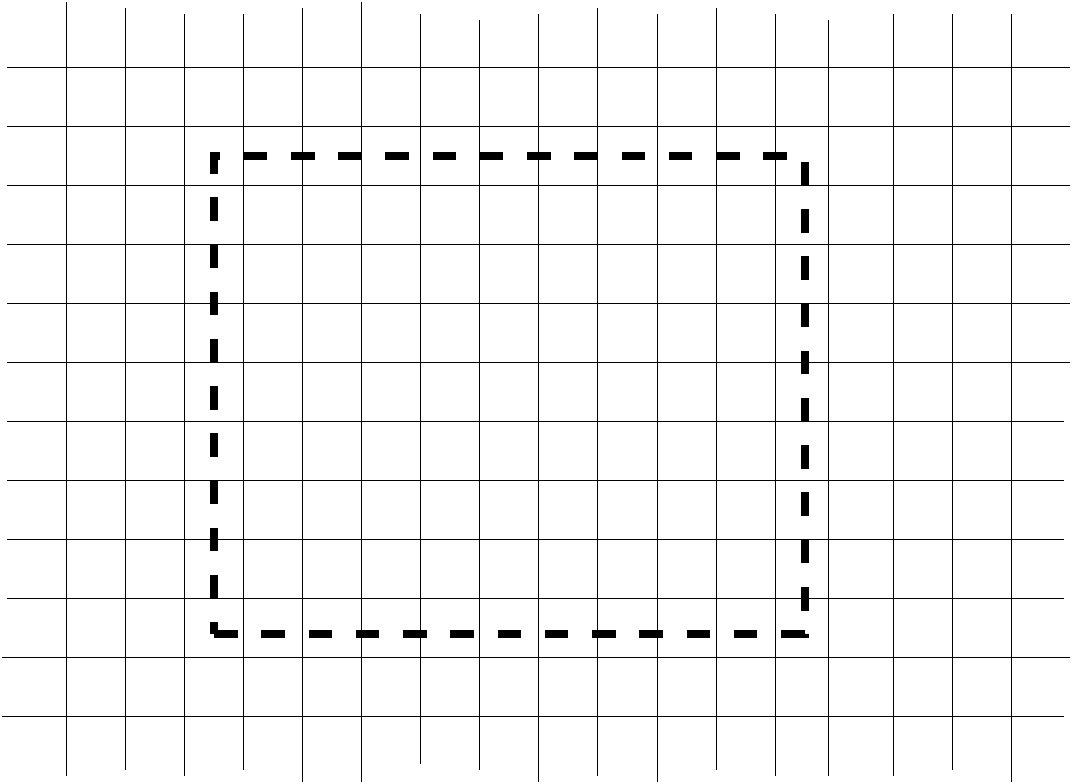}}
\caption{Control volumes for the continuous balance laws (dashed) and for particle averages (solid).}\label{fig:volume}
\end{figure}

The defining feature of the Eulerian flow model is that it is derived entirely in an Eulerian frame and it takes the microscopic granularity of a gas into account. In \cite{Svard18}, the process of going from a particle cloud to a continuum and on to a flow model was discussed at length. The key ideas can be explained using figure \ref{fig:volume}. The small boxes (solid lines) are taken to be larger than the mean free path, but only so much that they contain a sufficient number of molecules to produce sensible averages for the macroscopic variables $\rho,u,v,w,E,p,T$. It is also assumed that the gas is sufficiently dense such that thermodynamic equilibrium settles very quickly in each box. (These are the same assumptions as for the NSF system.) Associating each average with the centre point of its box, gives a grid of points where the macroscopic variables are known. By extrapolating from these points to the whole space (in a conservative way), the continuous macroscopic variables are formed. As a consequence, a point value in the continuum represents the average in a very small, but most importantly \emph{finite} box\footnote{Note the difference from kinetic theory where a particle distribution function, representing a large number of infinitesimally small molecules, is defined at each point in $\Rb^3$}. Having constructed the continuum variables, the flow model can be derived. In the Eulerian frame this is done by considering the dashed control volume, which has to be considerably larger than the smaller boxes, and calculating the convective and diffusive flows through its boundaries. The convective part is identical to the compressible Euler equations and for the diffusive part, one notes that the random motions of molecules in the vicinity of the (dashed) control-volume boundary will randomly transport molecules across the boundary. When a molecule passes the boundary its mass, momentum and (kinetic) energy follows, which is modelled as the diffusion terms scaled with a single $\nu$ in (\ref{eulerian}). That is, $\nu$ represents the rate at which molecules pass the dashed line and this rate is clearly the same for all conserved variables\footnote{The kinetic energy at the molecular level is represented by the total energy, $E$, at the continuum level.}.

However, there is another dissipative mechanism that is not accounted for in the derivation of the original model. The macroscopic variables are generally not constant within the (dashed) control volume and the relaxation acts on the length scale of the mean free path, which is much smaller than the (dashed) control volume. Modulo the transport across the boundary, $\rho, \rho\uu, E$ are all conserved since collisions within the dashed control volume do not affect the means of the conserved variables in the control volume. However, collisions may affect the balance between the internal and kinetic energy. The temperature is proportional to the random part of the velocity of gas particles within one of the small boxes in fig. \ref{fig:volume}. By diffusion, and subsequent collisions, the temperature will tend to even out between the small boxes \footnote{ Since gas molecules move in relation to each other before they collide, we label the process as \emph{diffusive} rather than conductive. Presumably, it is called conductive in the NSF system since molecules are not ``allowed'' to diffuse in a mass-element view since that in turn would imply mass diffusion.}. The only entropy consistent dissipative term that can be added to the energy equation to account for intra-control volume relaxation is a heat diffusive term of the form 
\begin{align}
(\kappa_TT_{x})_x.
\end{align}
We remark that the extra diffusive term is a consequence of the different scales that appear in the transition from a particle cloud to a continuum model. The continuum model can \emph{not} be derived at the continuum scale (small boxes in fig. \ref{fig:volume}), which opens for diffusion within the (dashed) control volume. Moreover, this means that one can not regard fluid elements, be they volume or mass elements, as infinitesimal. Despite the PDE formulation, the modelling error is still at the (dashed) control volume scale. (Of course, the same is true for the NSF.)


\section{Measuring diffusive heat transfer in gases}\label{sec:hot_wire}




There are a few different ways to experimentally determine the thermal conductivity. Here, we consider the transient hot-wire technique. In this technique, the gas is kept at constant pressure, a wire is heated and the heat dissipation is measured before convection becomes a significant factor. The experimental data is subsequently compared with solutions to the heat equation
\begin{align}
c_p\rho T_t =\kappa T_{xx}.\label{heat_ref}
\end{align}
The thermal conductivity $\kappa$ is determined to obtain the best match between (\ref{heat_ref}) and the measured data. If done carefully, which includes an accurate numerical solution procedure for (\ref{heat_ref}), this leads to errors in $\kappa$ of less than $0.2-0.5\%$ for gases according to \cite{Assael_etal23}. 

We remark that tabulated data for $\kappa$ need not be quite as accurate as $0.2-0.5\%$. Nevertheless, by designing the flow model to match (\ref{heat_ref}) any improvements in $\kappa$ for a particular gas would also improve the flow model.

However, since $\kappa$ is not obtained by directly matching the NSF system to experimental data, there \emph{may} be  a  mismatch between (\ref{heat_ref}) and NSF system. We will quantify that difference below.

\section{Diffusive heat transfer in the flow models}\label{sec:heat_cond}

To study heat diffusion in the flow models, it suffices to consider the 1-D versions of the equations on the domain $\Omega=[0,L]$ with periodic boundary conditions, and $t\in [0,\T]$. 
The 1-D Navier-Stokes-Fourier equations are,
\begin{align}
  \rho_t + (\rho u)_x &= 0, \nonumber\\
  (\rho u)_t + (\rho u^2+p)_x &= \frac{4}{3}\mu u_{xx}+\zeta u_{xx}, \label{1D-NS}\\
  E_t + (u(E+p))_x &= \frac{4}{3}\mu (uu_{x})_x +\zeta (uu_{x})_x+\kappa T_{xx}. \nonumber
\end{align}
These equations are closed with the ideal gas law, $p=\rho R T$. The dynamic viscosity coefficient $\mu$ and the thermal conductivity $\kappa$ are related via the Prandtl number, $Pr=\frac{c_p\mu}{\kappa}$. $\zeta$ denotes the bulk viscosity. 

In the hot-wire experiments, $u=0$, or at least very close to zero. The pressure, $p=p_0$, is constant and there is temperature gradient. Since $p=p_0$, we have $p_x=0$, implying that the momentum equation reduces to $(\rho u)_t=0$. Furthermore, with $u=0$, the continuity equation reduces to $\rho_t=0$. Thus, $\rho=\rho(x)$. Turning to the energy equation, we obtain under the same assumptions,
\begin{align*}
(\rho c_v T)_t =\kappa T_{xx}.
\end{align*}
Since $\rho_t=0$, we have
\begin{align*}
T_t =\frac{\kappa}{c_v \rho} T_{xx}.
\end{align*}
This resembles (\ref{heat_ref}), but can be recast into an equation in $T$ alone.
To this end, we use the gas law and obtain $T(x)=\frac{p_0}{R\rho(x)}$.  Using this relation, we arrive at
\begin{align}
T_t =\frac{\kappa R T(x)}{c_v p_0} T_{xx},\label{not-Fourier}
\end{align}
implying that temperature must be time-dependent, which is a contradiction.  It means that the assumptions, $p_0$ and $u=0$ do not lead to a solution to the Navier-Stokes-Fourier equations. In other words, determining $\kappa$ using (\ref{heat_ref}) as a model is not the same as using (\ref{1D-NS}). (Note that this has nothing to do with measurement errors.) This is of course well-known; a heat gradient will induce a velocity, which is why in the \emph{transient hot-wire technique} measurements must be conducted before convection is significant such that (\ref{heat_ref}) is an accurate model. It is mentioned in \cite{Assael_etal23} that measurements need to be carried out within one second to ensure this. We will confirm this in the numerical experiments below.

\vspace{0.5cm}

Turning to the Eulerian flow model, the 1-D equations are:
\begin{align}
  \rho_t + (\rho u)_x &= (\nu \rho_x)_x, \nonumber \\
  (\rho u)_t + (\rho u^2+p)_x &= (\nu (\rho u)_{x})_x,\label{1D-EFM} \\
  E_t + (u(E+p))_x &=  (\nu (E_{x}))_x+(\kappa_TT_{x})_x, \nonumber
\end{align}
where $\nu=\mu/\rho$ and $\kappa_T=0$ in the original model. We use the same assumptions, $p=p_0$ and $u=0$, to conclude that the momentum equation vanishes. The other two equations reduce to,
\begin{align*}
  \rho_t  &= (\mu (\ln\rho)_x)_x, \\
  (\rho c_v T)_t  &=  \left(\frac{\mu}{\rho}(\rho c_v T)_x\right)_x. 
\end{align*}
Recasting the energy equation
and using the continuity  equation in the energy equation yields,
\begin{align*}
  \rho c_v T_t  &= -c_vT\mu (\ln\rho)_{xx} + \mu\left(c_v T(\ln \rho)_x +  c_v T_x\right)_x,
\end{align*}
or 
\begin{align*}
  \rho c_v T_t  &=  \mu c_v T_{xx} + \mu c_vT_x(\ln\rho)_x. \nonumber
\end{align*}
From the gas law and $p_x=0$, we have  $\frac{T_x}{T}=-\frac{\rho_x}{\rho}$, which we use to obtain
\begin{align*}
  T_t  &=  \frac{\mu}{\rho} T_{xx} +\mu \frac{T_x}{\rho}(\ln\rho)_x  =  \frac{\mu}{\rho} T_{xx} -\frac{\mu}{\rho} \frac{T_x^2}{T}.
\end{align*}
Since we are considering hot-wire experiments at normal temperatures, we can assume that the first term on the right-hand side dominates the second and arrive at the following approximate relation
\begin{align}
  T_t  &\approx  \frac{\mu}{\rho} T_{xx}  = \frac{\mu RT}{p_0}T_{xx} ,\label{approx_eulerian}
\end{align}


Next, we calculate the discrepancy  between the heat diffusion coefficients of the EFM (\ref{approx_eulerian}) and the heat equation (\ref{heat_ref}). 
\begin{align*}
\frac{\kappa}{\rho c_p} - \frac{\mu}{\rho}=\frac{\kappa}{\rho c_p} - \frac{Pr\kappa}{c_p\rho}=\frac{\kappa}{\rho c_p}(1-Pr).
\end{align*}
This suggests that we should take $\kappa_T=\kappa_E$ given by
\begin{align}
\kappa_E=\kappa(1-Pr)=\frac{\mu c_p}{Pr}(1-Pr),\label{kappa_E}
\end{align}
in (\ref{eulerian})/(\ref{1D-EFM}).

We end this section with the observation that adding a heat diffusive term to the energy equation implies that there is no mode of pure thermodynamic relaxation in the Eulerian flow model. It was postulated in \cite{Svard18} that this is a desirable feature since otherwise a velocity will be induced at scales where diffusion should dominate.  With $\kappa_T>0$, a velocity will be induced although its magnitude is much smaller than that in the NSF as will be evident in the simulations below (see figure \ref{fig:extra}). At a global scale,  varying $T$ and $\rho$ will induce a velocity. Hence, it appears that the postulate can not be strictly enforced in a continuum model. There is, however, a difference to the NSF. In NSF a non-zero velocity is necessary to transport mass and effectuate the relaxation. For EFM, the non-zero velocity is simply induced as a consequence of the relaxation of $\rho$ and $T$.

\section{Numerical verifications}\label{sec:num_verif}

We are not directly trying to mimic the setup in an hot-wire experiment. Instead, we use solutions to the heat equation (\ref{heat_ref}) as the correct reference, since it has been matched to experiments.

To verify the theoretical derivations, we approximated (\ref{heat_ref}), (\ref{1D-EFM}) and  (\ref{1D-NS})  with periodic boundary conditions, by central finite volumes in space and a 4th-order Runge-Kutta method in time. The schemes have no built-in/explicit artificial diffusion that could distort the diffusive properties of the numerical solutions, and they have been implemented in Julia, \cite{julia}.

We consider a 1 meter long domain $x\in (0,1]$ and a relatively small sinusoidal perturbation in the temperature. We discretise the domain with N equidistant grid points.  The grid spacing is denoted  $\Delta x$. The finest grids we used have $N=400$, which ensures that the solutions are very resolved. 

The initial data are
\begin{align}
  \rho=1,\quad   u=0,\quad 
  T=273.0 + \delta\sin(2\pi x),\quad \delta =3.0.\nonumber
\end{align}
We use the following physical parameters for air 
\begin{align*}
  \mu = 18.1\cdot 10^{-6},\quad
  \gamma =1.4,\quad
  c_v=718,\quad
  Pr=0.71,
\end{align*}
and argon,
\begin{align*}
  \mu = 20.64\cdot 10^{-6},\quad
  \gamma = 1.661,\quad
  c_v=313,\quad
  Pr = 0.661.
\end{align*}
The remaining constants are given by
\begin{align*}
  c_p = \gamma c_v,\quad 
  \kappa = \frac{c_p\mu}{Pr}, \quad
  R=c_p-c_v.\nonumber
\end{align*}
We run the simulations with a time step $\Delta t =\frac{CFL}{c_0}\Delta x$, where $ CFL=1$ and $c_0$ is the speed of sound  at $t=0$ and for the mean temperature.  With $N=400$ and for air, 132193 time steps are required to march til $\T=1$.

\vspace{0.1cm}

\begin{table}[ht]
\centering
\begin{tabular}{|c | c | c | c | c|} 
 \hline
$\max|T^N-T^E|$ & $\max|T^N-T^H|$ & $\max|T^E-T^H|$ & $\max|u^N|$ & $\max|u^E|$ \\ 
 \hline
 8.67e-4 & 1.15e-5 & 8.97e-4 & 2.22e-6 & 0.0\\
 \hline
\end{tabular}
\caption{Temperature differences between the (original) EFM (\ref{1D-EFM}) with $\kappa_T=0$ (superscript E), NSF (\ref{1D-NS}) (superscript N), and heat equation (\ref{heat_ref}) (superscript H), for air and at time $\T=1$s.}
\label{table:orig}
\end{table}



\begin{figure}[ht]
  \centering
  \subfigure[N=400]{\includegraphics[width=8cm]{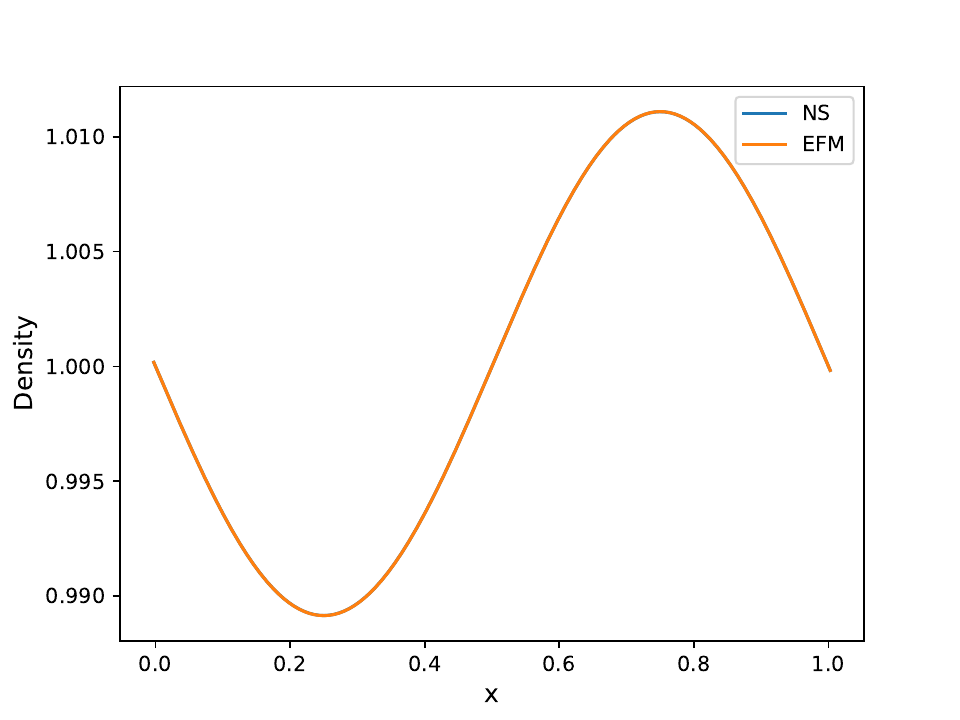}}
    \subfigure[N=400]{\includegraphics[width=9cm]{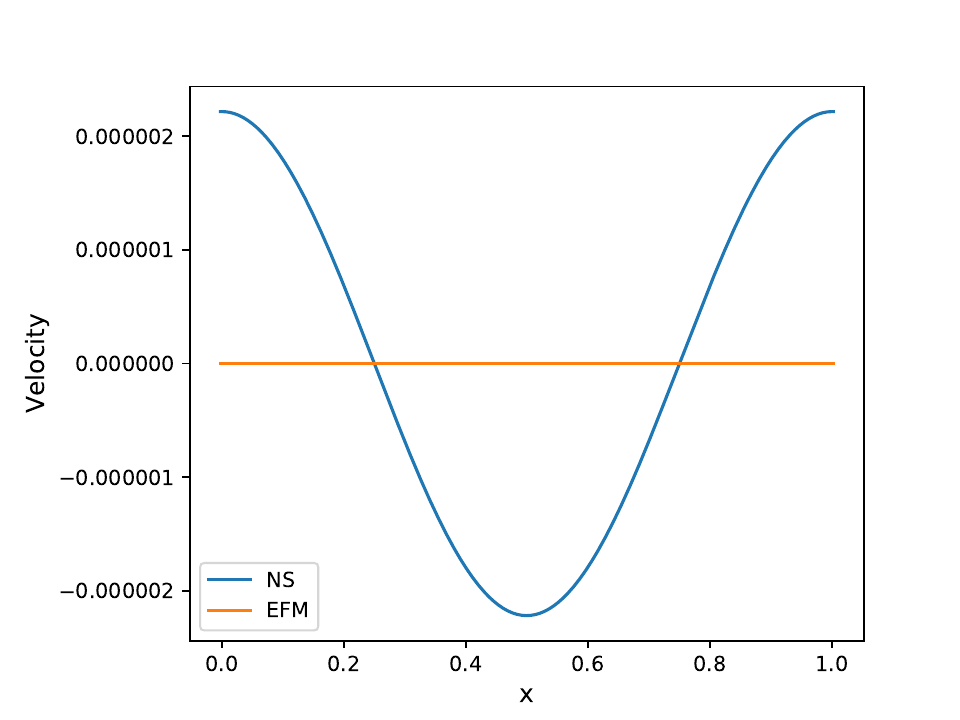}}
\caption{Results for air for the original EFM (\ref{1D-EFM}) ($\kappa_T=0$) and the NSF (\ref{1D-NS}) at $\T=1$s. }\label{fig:orig1}
\end{figure}

\begin{figure}[ht]
  \centering
    \subfigure[N=400]{\includegraphics[width=8cm]{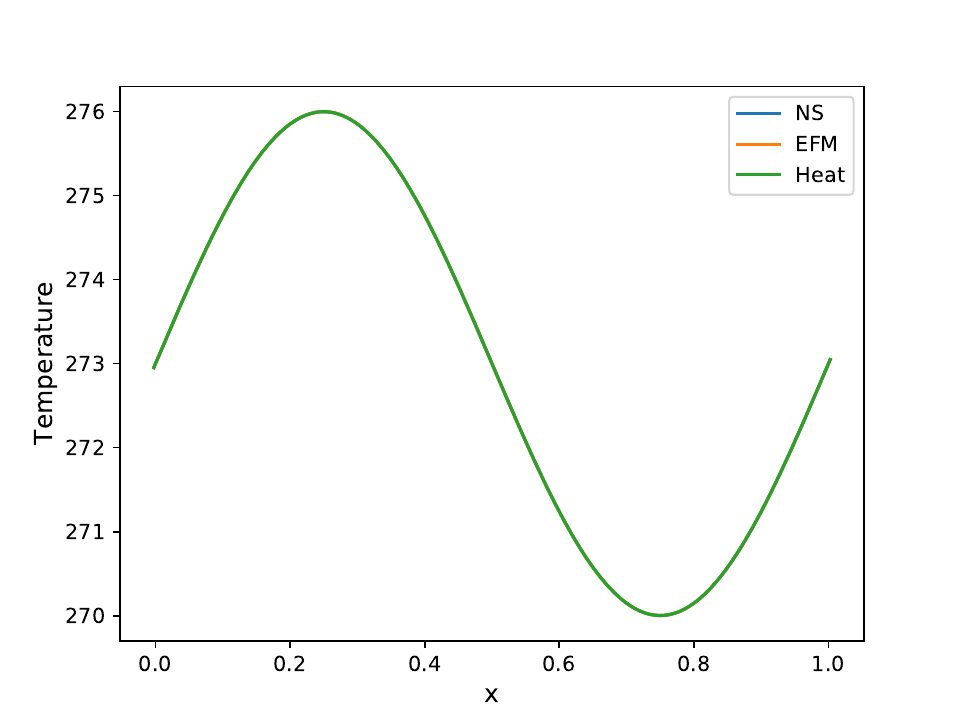}}
    \subfigure[N=400]{\includegraphics[width=8cm]{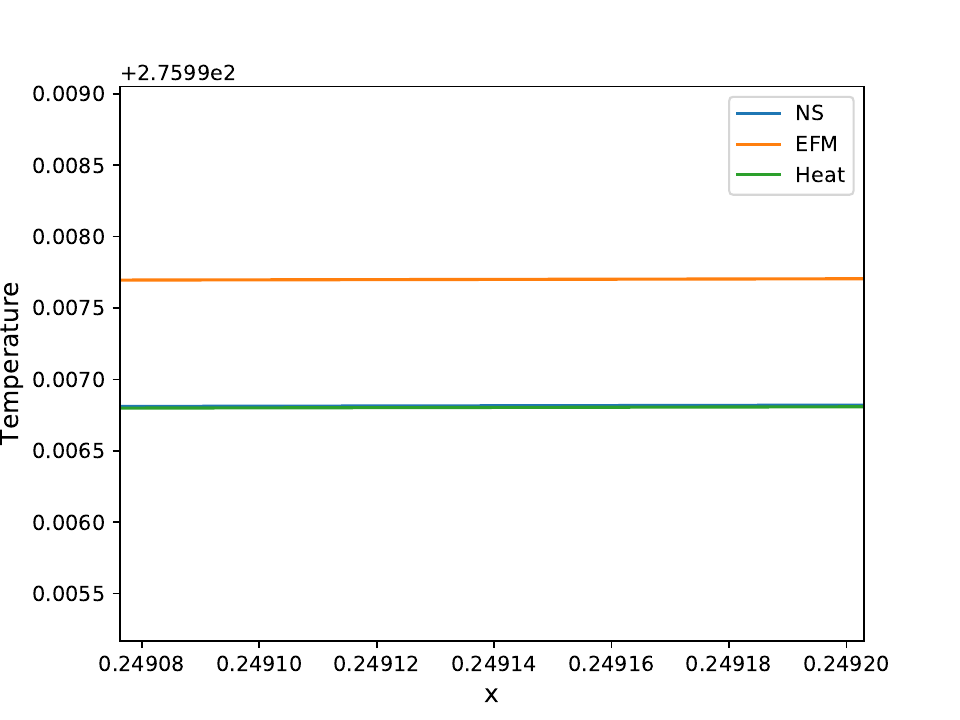}}
\caption{Results for air for the original EFM (\ref{1D-EFM}) ($\kappa_T=0$), the NSF (\ref{1D-NS}) and the heat equation, at $\T=1$s.}\label{fig:orig2}
\end{figure}
We begin with the results for air, and without the extra heat diffusive term, i.e., $\kappa_T=0$. From table \ref{table:orig} and figures \ref{fig:orig1} and \ref{fig:orig2}, we see that the magnitude of the velocity after 1 second is indeed minuscule. Furthermore, we see that the temperature difference between the heat equation and the Navier-Stokes-Fourier solution is of the order $10^{-5}$.
\begin{remark}
The numerical errors can be estimated by comparing solutions with $N=200$ and $N=400$. These differences are three orders of magnitude smaller than the differences between the models. This ensures that the differences we see between the models are actually due to the models and not numerical errors.
\end{remark}
Turning to the EFM, we first note that no velocity is induced as expected. However, the temperature difference to the heat equation is almost two orders of magnitude larger than the difference between the heat equation and the NSF system. Furthermore, the differences are not the result of uncertainties in $\kappa$, that are of the order a few per cent. 

Nevertheless, the differences between the Eulerian flow model and the Navier-Stokes-Fourier are very, very small in absolute terms. In the unzoomed figures all three solutions lie on top of each other. It is only in the zoom in figure \ref{fig:orig2} that the difference can be seen. This may explain why the two systems produced almost identical results for aerodynamic applications in \cite{SayyariDalcin21,SayyariDalcin21_2,DolejsiSvard21}. The new heat diffusive term is in these applications insignificant compared to the main diffusive terms and convection.

\begin{figure}[ht]
  \centering
    \subfigure[N=400]{\includegraphics[width=8cm]{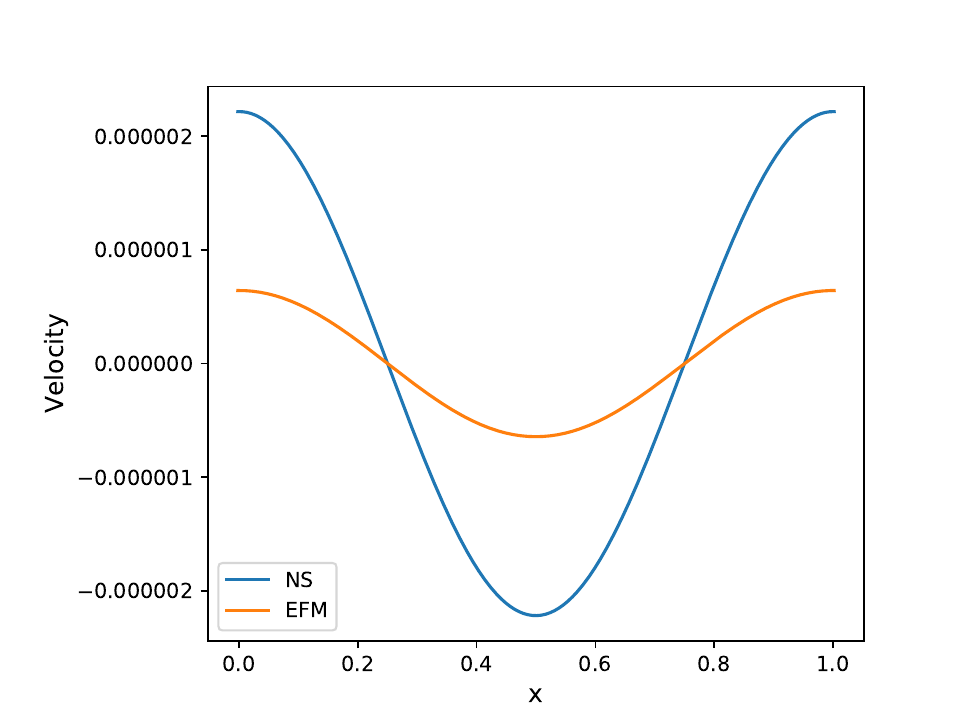}}
    \subfigure[N=400, zoom of temperature]{\includegraphics[width=8cm]{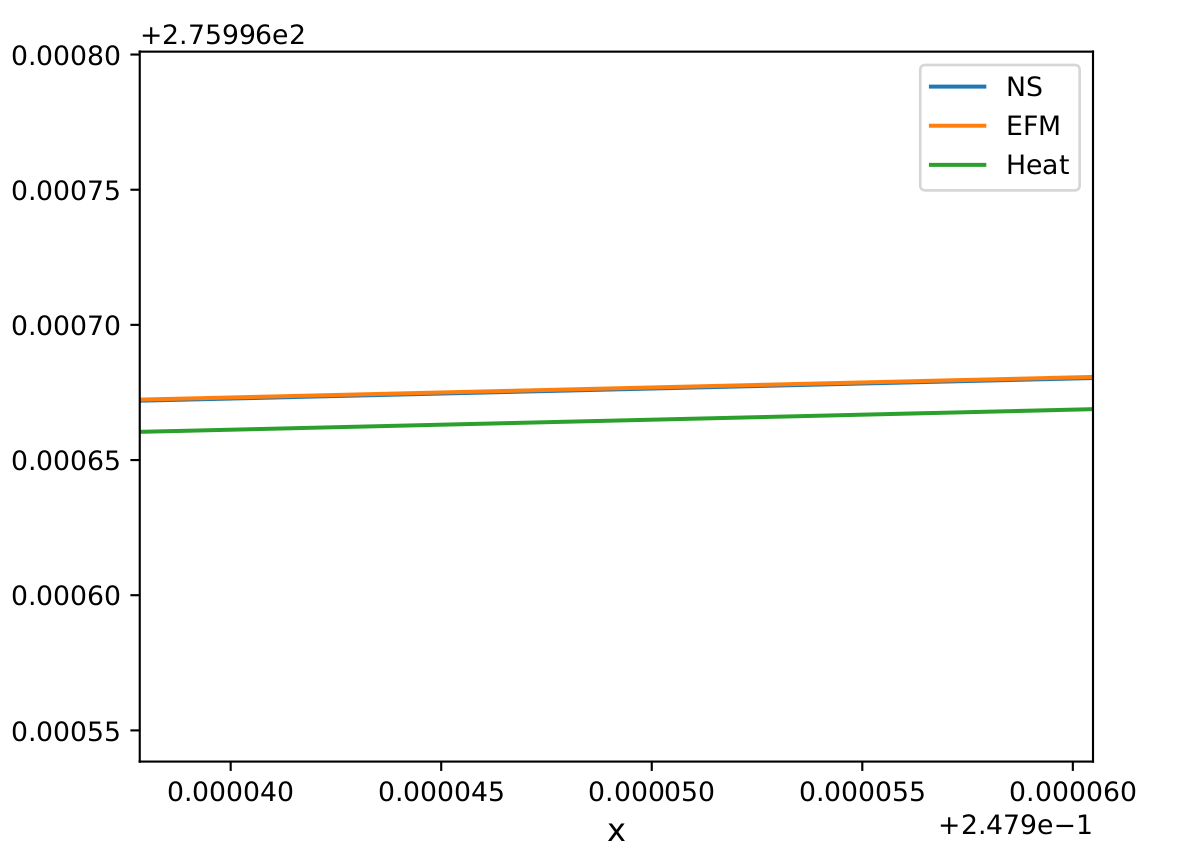}}
\caption{Results for EFM (\ref{1D-EFM}) with (\ref{kappa_E}),  NSF (\ref{1D-NS}), and heat equation (\ref{heat_ref}), for air and $\T=1$s.}\label{fig:extra}
\end{figure}

Next, we repeat the numerical experiments and include the heat diffusive term (\ref{kappa_E}) in  (\ref{1D-EFM}). The results are displayed in table \ref{table:extra} and figure \ref{fig:extra}. In this case the difference between NSF and EFM is two orders of magnitude less than their differences to the heat equation.  
\begin{table}[ht]
\centering
\begin{tabular}{|c | c | c | c | c|} 
 \hline
$\max|T^N-T^E|$ & $\max|T^N-T^H|$ & $\max|T^E-T^H|$ & $\max|u^N|$ & $\max|u^E|$ \\ 
 \hline
 3.93e-7  & 1.15e-5 & 1.19e-5 &  2.22e-6 & 6.43e-7 \\
 \hline
\end{tabular}
\caption{Temperature differences between the EFM (\ref{1D-EFM}) (superscript E) with $\kappa_T$ given by (\ref{kappa_E}), the NSF (\ref{1D-NS}) (superscript N), and heat equation (\ref{heat_ref}) (superscript H), for air and at $\T=1$s. }
\label{table:extra}
\end{table}

Furthermore, with the extra diffusion term, the Eulerian model induces a velocity, albeit much smaller than the Navier-Stokes-Fourier, figure \ref{fig:extra}. \emph{The larger velocity in the NSF solution accounts for the transport of internal energy that is effected by the mass diffusion in the EFM.}

Finally, we repeat the same numerical experiments for argon. The results are very similar and we only list the corresponding differences in table \ref{table:argon} and \ref{table:argon_extra}. With $\kappa_T=\kappa_E$ given by (\ref{kappa_E}), the temperature difference between NSF and EFM effectively vanishes.
\begin{table}[ht]
\centering
\begin{tabular}{|c | c | c | c | c|} 
 \hline
$\max|T^N-T^E|$ & $\max|T^N-T^H|$ & $\max|T^E-T^H|$ & $\max|u^N|$ & $\max|u^E|$ \\ 
 \hline
1.24e-3  & 1.35e-5 & 1.23e-3  &  2.50e-6 & 0.0\\
 \hline
\end{tabular}
\caption{Temperature differences between the original EFM (\ref{1D-EFM}) with $\kappa_E=0$ (superscript E), NSF (\ref{1D-NS}) (superscript N), and heat equation (\ref{heat_ref}) (superscript H) for argon and $\T=1$s.}
\label{table:argon}
\end{table}




\begin{table}[ht]
\centering
\begin{tabular}{|c | c | c | c | c|} 
 \hline
$\max|T^N-T^E|$ & $\max|T^N-T^H|$ & $\max|T^E-T^H|$ & $\max|u^N|$ & $\max|u^E|$ \\ 
 \hline
8.20e-7  & 1.35e-5 & 1.44e-5 &  2.50e-6 & 8.47e-7\\
 \hline
\end{tabular}
\caption{Temperature differences between the EFM (\ref{1D-EFM}) with  $\kappa_E$ given by (\ref{kappa_E}) (E), the NSF (\ref{1D-NS}) (N) and heat equation (\ref{heat_ref}) (H), for argon and $\T=1$s.}
\label{table:argon_extra}
\end{table}



\noindent {\bf Modelling error: } In all simulations we see that the EFM, augmented with the heat diffusive term (\ref{kappa_E}), and NSF system are very close. Indeed, closer to each other than to the solution to the heat equation, which we have assumed track the experimental data. By rerunning the tests and slightly varying $\kappa$ in NSF and $\kappa_E$ in EFM, we can estimate the size of this error and we find that by increasing $\kappa$ approximately $0.5\%$ and $\kappa_E$ by roughly $1\%$ they match the heat equation closely. These modelling errors are of the same order as the optimal measurement errors for $\kappa$ and thus of little concern.


\vspace{0.1cm}

\noindent {\bf Experimental-like test case:} By reducing the domain length to 1 cm ($x\in [0,0.01]$) and taking a temperature variation of $\delta=1.0$ Kelvin, we obtain a setup that, with respect to scales, resemble the hot-wire experiments.

We run the code  for air with the extra heat diffusive term in EFM til $\T=1$ second with $N=100,200,400$. (The finest grid required 1.3 million time steps.) The maximal velocities for NSF and EFM decreased with finer grids and with $N=400$ they were $1.2e-7$ and $2.2e-8$ respectively, i.e. about 5 times larger for NSF.  Nevertheless, in both cases, the velocities after 1 second are minuscule, confirming the statement in  \cite{Assael_etal23}.

The maximal temperature difference between EFM and NSF also decreased as the grids were refined and with $N=400$ it was $1.6e-7$. The difference between EFM/NSF and the heat equation converged quickly and was $1.03e-3$ for all grids. Although this is a very small difference in absolute numbers, it is still several orders of magnitude larger than the EFM-NSF difference.


However, the gap between the heat equation and EFM is effectively closed by a $4\%$ increase of $\kappa_E$, and between the heat equation and for NSF with about $2\%$ increase of $\kappa$. Note that, the absolute changes in the diffusion coefficients are the same but, $\kappa_E$ needs a relatively larger change due to its smaller magnitude. One may also note that the correction is different from the first experiment; that is, $\kappa$ \emph{does not appear to be perfectly scale independent}.





\section{Sound attenuation}\label{sec:sound}

In \cite{SvardMunthe23}, the attenuation  for small (linear) amplitude sound waves, was studied for the Eulerian flow model (\ref{eulerian}). The \emph{attenuation coefficients}  (see \cite{SvardMunthe23}) for the NSF system (\ref{1D-NS}) and the Eulerian system (\ref{1D-EFM}) are given by
\begin{align}
  \Gamma_{NSF}&=\frac{\omega^2}{2c^2_0\rho}\mu(\frac{4}{3}+\frac{\zeta}{\mu}+\frac{\gamma-1}{Pr}),\label{GNSF}\\
  \Gamma_{EFM}&=\frac{\omega^2}{2c^2_0\rho}\mu(2+\cf(1-Pr)\frac{\gamma-1}{Pr}),\nonumber
\end{align}
where $\omega=2\pi f$ is the angular frequency; $f$ the frequency; $c_0$ the speed of sound; $\zeta$ the bulk viscosity. The parameter $\cf$ takes the value $\cf=1$, if the correction is included, i.e., $\kappa_T=\kappa_E$ given by (\ref{kappa_E}), and $\cf=0$ if $\kappa_T=0$.

For monatomic gases, it is generally accepted that the bulk viscosity $\zeta=0$, although this has been questioned in \cite{Rajagopal13}. However, according to \cite{Buresti15}, the experimental evidence is scarce but indicates that the bulk viscosity for monatomic gases is indeed practically negligible. Therefore, to take $\zeta=0$, $\gamma=5/3$ and $Pr=2/3$ seems reasonable for monatomic gases. Then with $\cf=0$, as in the original EFM,  $\Gamma_{NSF}\neq \Gamma_{EFM}$ but experimental data for argon in \cite{Greenspan56} favour the standard model (\ref{1D-NS}).  However, the experimental errors were not disclosed, it was therefore left open in \cite{SvardMunthe23}  whether or not the Eulerian model (\ref{eulerian}) needed a modification.  Nevertheless,  for ideal monatomic gases ($Pr=2/3$ $\gamma=5/3$ and $\zeta=0$)  a correction to (\ref{eulerian})  was calculated in \cite{SvardMunthe23}. That correction corresponds to (\ref{GNSF}) with $\cf=1$, in which case $\Gamma_{NSF}=\Gamma_{EFM}$. That is, with the correction, (\ref{eulerian}) matches the experimental data for argon. (Although, we still do not know the experimental accuracy.)




Turning to diatomic Newtonian gases, it is routinely assumed that Stoke's hypothesis, asserting that $\zeta=0$, is valid. If so, then with $Pr=0.71$, $\gamma=1.4$ and  $\cf=0$,  (\ref{GNSF}) leads to $\Gamma_{EFM}>\Gamma_{NSF}$.


However, the difference $\Gamma_{EFM}-\Gamma_{NSF}$ is small. (About $6\%$.) In \cite{Ejakov03}, sound attenuation experiments with diatomic gases were conducted and the uncertainties exceed any discrepancies between $\Gamma_{EFM}$ and $\Gamma_{NSF}$. Adding the correction $\cf=1$ would increase the difference but still within experimental accuracy of the sound attenuation experiments.  Moreover, it was shown in  \cite{SvardMunthe23} that $\kappa_T>0$ improved the accuracy of shock profiles for nitrogen, which suggests that adding the new diffusive term improves the EFM.

Furthermore, Stokes' assumption, $\zeta=0$, is known to be incorrect for diatomic gases where  $\zeta\sim 1$ (\cite{Cramer12}), implying that $\Gamma_{NSF}$ is significantly larger.  In fact, ultrasound experiments (such as \cite{Ejakov03,Greenspan56}) is one of few  methods to estimate the bulk viscosity (\cite{Buresti15}) which means that $\mu$ and $\kappa$ have to be known \emph{a priori}. For instance, from hot-wire experiments. In summary, the sound attenuation coefficient for diatomic gases is larger than $\Gamma_{NSF}$ with $\zeta=0$. This also supports adding the new diffusive term to the EFM, but since $\zeta$ is largely unknown, accurate comparisons between NSF and EFM are currently difficult. 

In practice,  variations in humidity or pollutants in the gas have a far greater impact than rather substantial uncertainties in $\zeta$ (or $\kappa_T$). However, for its intended applications, the NSF system produces as accurate results as one can expect from a continuum model. The same is true for the EFM.

\section{Vortex simulation}\label{sec:vortex}


Next, we turn to the effect of of $\kappa_E$ on aerodynamic applications where the EFM has been shown to yield very accurate results with $\kappa_T=0$.  It is beyond the scope of this paper to repeat all the simulations run in \cite{SayyariDalcin21,SayyariDalcin21_2,DolejsiSvard21}. Here, we study a commonly used smooth vortical flow that we place in a  2-D box with solid wall boundaries. Since there is no background flow, it will remain stationary and slowly decay. This is possible to resolve to a very high accuracy even on a desktop computer. 

The domain is $\Omega=[0,1]^2$ and we impose wall boundary conditions at all boundaries:
\begin{align}
\frac{\partial\rho}{\partial n}=0, \quad \frac{\partial T}{\partial n}=0, \quad (u,v)=(0,0).
\end{align}
The domain is discretised with $N^2$ equidistant grids points ($N=\{1000,500\}$). We run the convergent second-order accurate scheme given in \cite{Svard22}, discretised with a third-order strong stability preserving Runge-Kutta scheme in time; the scheme is implemented in FORTRAN with MPI parallelisation. 

The vortex has initial conditions is given by
\begin{align*}
  f (x,y)   & = \frac{ (x-x_0)^2 + (y-y_0)^2}{\RR^2 } \\
  u(x,y)    & = -\beta \frac{y-y_0}{\RR} \exp(-\frac{f}{2})\\
  v(x,y)    & =  \beta \frac{x-x_0}{\RR} \exp(-\frac{f}{2})\\
  T(x,y)    & = T_\infty- \frac{1}{2c_p} \beta^2 \exp(-f)\\
  \rho(x,y) & = \rho_\infty \left(\frac{T(x,y)}{T_{\infty}}\right)^{\frac{1} { \gamma-1}}
\end{align*}
    where 
\begin{align*}
  x_0=0.5,\quad  y_0=0.5,\quad     \RR=0.1,\quad    \beta = 33.0,\quad  \\
  T_\infty=273.15,\quad
  \rho_\infty=1, \quad c_p=1005, \quad \gamma=1.4,\quad R=287.15.
\end{align*}
which gives a vortex with radius $0.1$ centred in the box. Its maximal initial velocity is about 20 m/s. The physical constants correspond to air. Moreover, we have used $\mu=1e-3$, which gives a radius-based Reynolds number of 2000.


First, we ran the simulations with and without $\kappa_E$ on the $N=1000$ grid til $\T=0.005$, which corresponds to 0.16 revolutions. We recorded the maximal differences,
\begin{align*}
 \|\rho_\kappa-\rho\|_{\infty}& =    4.0e-6\\
 \|(\rho u)_\kappa-(\rho u)\|_{\infty}& =    7.6e-5\\
 \|(\rho v)_\kappa-(\rho v)\|_{\infty}& =    7.5e-5\\
 \|E_\kappa-E\|_{\infty} &=       1.0e-2
\end{align*}
where the subscript $\kappa$ signifies the numerical solution with the extra diffusive term. Despite the rather different sizes of the maximal errors, these all correspond to \emph{relative errors} of less than $1e-6$, which amounts to very small differences. Of course, the small differences are partly a consequence of the short run. However, a more immediate problem is that of numerical errors. Running the same problem on the $N=500$ grid and comparing the numerical solutions (without the extra diffusion) on the two grids, reveal that the differences are of the same orders as the differences between the two models. The conclusion is that the differences for the EFM with $\kappa_E$ given by (\ref{kappa_E}) and with $\kappa_E$ is less than the current numerical errors. That is, \emph{beyond the 7th accurate digit}.  

\begin{figure}[ht]
  \centering
  \subfigure[N=500]{\includegraphics[width=8cm]{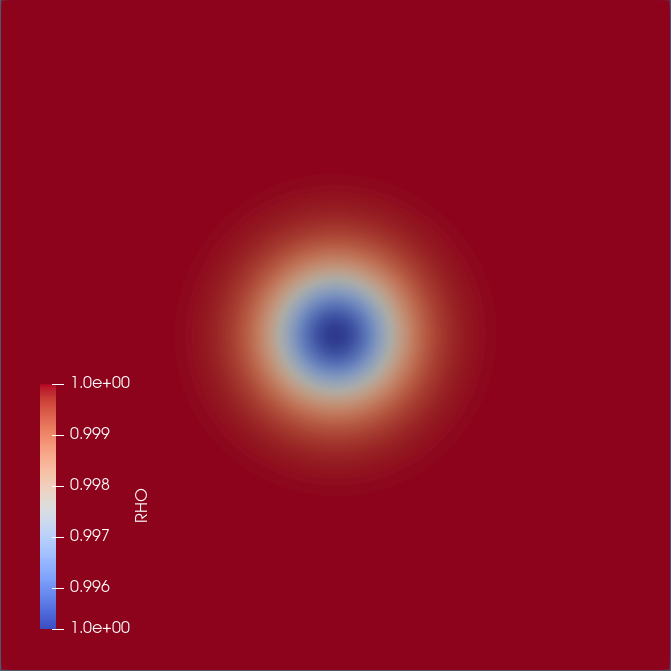}}
\caption{The vortex at $\T=0.05$ computed with $500^2$ grid and including the extra $\kappa_E$ term. }\label{fig:vortex}
\end{figure}

Next, we ran the simulations 10 times longer, til $\T=0.05$, i.e., 1.6 revolutions, and the $N=500$ grid. The numerical solution with the  $\kappa_E$-diffusion is displayed in fig. \ref{fig:vortex}.  In this case, the difference between the two versions of the EFM increased by less than a factor of 10, which is what one would expect merely from the accumulation of numerical errors. Even so, these simulations agree to within 6 accurate digits. 

Even $N=500$  must be considered a very high resolution for this problem and yet we can not quantify the differences between the versions due to lack of resolution. Hence, it is difficult to envisage that the addition of $\kappa_T=\kappa_E$ will have a significant effect for practical aerodynamic simulations.

\section{Conclusions}\label{sec:conclusions}

In the original model (\cite{Svard18}) $\kappa_T$ was not included. It was introduced in \cite{Svard22} in the form (\ref{kappar}) but  mainly as a technical assumption to aid in the proof of weak well-posedness. In view of the underlying molecular dynamics it appears that an extra diffusive term should be present in the energy equation of the EFM model. For approximately ideal gases, we have by theoretical and numerical analysis quantified it and concluded that 
\begin{align}
  \kappa_T=\kappa_E+4\kappa_rT^3\label{final_kappaT}
\end{align}
where
\begin{align}
\kappa_E=\kappa(1-Pr)=\frac{\mu c_p}{Pr}(1-Pr),\tag{\ref{kappa_E}}
\end{align}
  should be used in (\ref{eulerian}). This choice matches the heat diffusion of the NSF system and thus improves the heat transfer properties of the EFM; it improves shock profiles (see \cite{SvardMunthe23}), and it is very likely to improve sound attenuation properties. In view of the vortex results in section \ref{sec:vortex}, it is unlikely that the extra heat diffusive term resulting from $\kappa_E$ will have a measurable effect on the validation cases in (\cite{SayyariDalcin21,SayyariDalcin21_2,DolejsiSvard21}) (but that has to be verified).

We remark that high-frequency ultrasound waves and shock profiles are at the fringe of the validity range of these continuum models. Improvements in these regimes will have little practical impact since other errors dominate. Indeed, the NSF model is routinely used without accurate knowledge of the bulk viscosity. 

Finally, an important aspect that is often disregarded, is that the mathematical model used for practical predictions is solvable at all. This is where the NSF and the EFM differ. The proposed modification to (\ref{eulerian}),  \emph{is trivial to incorporate in the well-posedness results and in the convergent numerical scheme derived in \cite{Svard22}}. For the EFM model, it is certain that the numbers produced by the simulation code actually track the solution to the equations.


\begin{thebibliography}{AAVW23}

\bibitem[AAVW23]{Assael_etal23}
M.~J. Assael, K.~D. Antoniadis, D.~Velliadou, and W.~A. Wakeham.
\newblock Correct use of the transient hot‐wire technique for thermal
  conductivity measurements on fluids.
\newblock {\em International Journal of Thermophysics}, 44:1--19, 2023.

\bibitem[BKSE12]{julia}
Jeff Bezanson, Stefan Karpinski, Viral~B Shah, and Alan Edelman.
\newblock Julia: A fast dynamic language for technical computing.
\newblock {\em arXiv preprint arXiv:1209.5145}, 2012.

\bibitem[Bur15]{Buresti15}
Guido Buresti.
\newblock A note on {S}tokes' hypothesis.
\newblock {\em Acta Mech.}, 226:3555--3559, 2015.

\bibitem[Cra12]{Cramer12}
M.S. Cramer.
\newblock Numerical estimates for the bulk viscosity of ideal gases.
\newblock {\em Physics of fluids}, 24:066102--1--22, 2012.

\bibitem[DS21]{DolejsiSvard21}
V{\'i}t Dolej{\v s}{\'i} and Magnus Sv{\"a}rd.
\newblock Numerical study of two models for viscous compressible fluid flows.
\newblock {\em J. Comp. Phys}, 427:110068, 2021.

\bibitem[EPD{\etalchar{+}}03]{Ejakov03}
S.~G. Ejakov, S.~Phillips, Y.~Dain, R.~M. Lueptow, and J.~H. Visser.
\newblock Acoustic attenuation in gas mixtures with nitrogen: Experimental data
  and calculations.
\newblock {\em J. Acoust. Soc. Am.}, 113(4), 2003.

\bibitem[Gre56]{Greenspan56}
Martin Greenspan.
\newblock Propagation of sound in five monatomic gases.
\newblock {\em The Journal of the Acoustical Society of America},
  28(4):644--648, 1956.

\bibitem[KP22]{KajzerPozorski22}
Adam Kajzer and Jacek Pozorski.
\newblock The mass diffusive model of sv{\"a}rd simplified to simulate nearly
  incompressible flows.
\newblock {\em Computers \& Mathematics with applications}, 121:18--29, 2022.

\bibitem[Raj13]{Rajagopal13}
K.R. Rajagopal.
\newblock A new development and interpretation of the {N}avier-{S}tokes fluid
  whoch reveals that "{S}tokes assumption" is inapt.
\newblock {\em International Journal of Non-linear Mechanics}, 50:141--151,
  2013.

\bibitem[SDP21a]{SayyariDalcin21_2}
M.~Sayyari, L.~Dalcin, and M.~Parsani.
\newblock Development and analysis of entropy stable no-slip wall boundary
  conditions for the {E}ulerian model for viscous and heat conducting
  compressible flows.
\newblock {\em Partial Differential Equations and Applications}, 2:1:77--27:77,
  2021.

\bibitem[SDP21b]{SayyariDalcin21}
M.~Sayyari, L.~Dalcin, and M.~Parsani.
\newblock Entropy stable no-slip wall boundary conditions for the {E}ulerian
  model for viscous and heat conducting compressible flows.
\newblock In {\em AIAA-paper AIAA 2021-1662, AIAA Scitech 2021 Forum"}, 2021.

\bibitem[SM23]{SvardMunthe23}
M.~Sv\"ard and K.~Munthe.
\newblock A study of the diffusive properties of a modified compressible
  {N}avier-{S}tokes model.
\newblock {\em Meccanica}, 58:1083--1097, 2023.

\bibitem[Sv{\"a}18]{Svard18}
M.~Sv{\"a}rd.
\newblock A new {E}ulerian model for viscous and heat conducting compressible
  flows.
\newblock {\em Physica A}, 506:350--375, 2018.

\bibitem[Sv{\"a}22]{Svard22}
M.~Sv{\"a}rd.
\newblock Analysis of an alternative {N}avier-{S}tokes system: Weak entropy
  solutions and a convergent numerical scheme.
\newblock {\em M3AS}, 32(13):2601--2671, 2022.

\end{thebibliography}
\newcommand{\etalchar}[1]{$^{#1}$}



\end{document}